\documentclass[twocolumn,showpacs,preprintnumbers,amsmath,amssymb]{revtex4}

\usepackage{graphicx}
\usepackage{dcolumn}
\usepackage{bm}

\begin{document}

\title{Phase separation and vortex states in binary mixture of
Bose-Einstein condensates in the trapping potentials with displaced
centers}

\author{S. T. Chui}
\affiliation{Bartol Research Institute, University of Delaware, Newark, DE
19716}
\author{V. N. Ryzhov, and E. E. Tareyeva}
\affiliation{Institute for High Pressure Physics, Russian Academy of
Sciences, 142 190 Troitsk, Moscow region, Russia}

\date{24 January 2002}{*}

\begin{abstract}
The system of two simultaneously
trapped codensates consisting of $^{87}Rb$ atoms in two different
hyperfine states is investigated theoretically in the case when the minima of
the trapping potentials are displaced with respect to each other.
It is shown that the small shift of the minima
of the trapping potentials leads to the
considerable displacement of the centers of mass of the
condensates, in agreement with the experiment.
It is also
shown that the critical angular velocities of the vortex states of the system
drastically depend on the shift and
the relative number of particles in the condensates, and
there is a possibility
to exchange the vortex states between condensates by shifting the
centers of the trapping potentials.
\end{abstract}

\pacs{03.75.Fi,67.57.Fg,67.90.+z}

\maketitle

The experimental realization of Bose-Einstein Condensation (BEC)
in trapped alkali-atoms gases at ultralow temperatures offers new
opportunities for studying quantum degenerate fluids
\cite{[1],[2],[3]}. The art of manipulating these condensates,
which contain thousands of atoms confined to microscale clouds,
achieved very high level now. Creation of vortices in one
\cite{stir2000} and two-component BEC \cite{bose_vor,wilhol}
of $^{87}Rb$ atoms is the amazing example of this art.

In this article we focus on the properties of the
two-component BEC in the trap in which the trapping potentials for
each component are displaced with respect to each other in the
vertical direction \cite{dhall}. The condensate consists of the
simultaneously trapped otherwise identical atoms of $^{87}Rb$ in
two different hyperfine spin states $|1>$ and $|2>$ ($|1>$
and $|2>$ denote the $|F=1,m_f=-1>$ and $|2,1>$ spin states of
$^{87}Rb$ atoms respectively) \cite{dhall,twocom1,twocom2}.
The scattering lengths of the states $|1>$ and $|2>$ are known
to be in proportion $a_{11}:a_{12}:a_{22}=1.03:1.0:0.97$
with the average of the three being $55(3)\AA$
\cite{dhall,twocom1,twocom2}.

The double condensate system was prepared from the single $|1>$
condensate by driving a two-photon transition which transfers any
desired fraction of the atoms to the $|2>$ state by selecting
the length and amplitude of the two-photon pulse \cite{twocom1}.
The rotating magnetic field of the time-averaged orbiting potential
(TOP) trap gave the possibility to displace the minima of the trapping
potentials $V_1$ and $V_2$ with respect to each other. When the minima
of the trapping potentials were not shifted, the $|1>$ atoms formed
a shell about the $|2>$ atoms \cite{dhall}. This case has been
discussed theoretically in \cite{crt1,crt2}. If the minima of the
trapping potentials $V_1$ and $V_2$ are displaced from each other by
a distance which is small compared to the size of the total condensate
the resulting separation of the centers of mass of the condensates
is much larger \cite{dhall}. In this paper we provide for an
analytical explanation of this result.

In order to explore the boundary between the two condensates, we
begin with the analysis of their behaviour in the framework of the
Thomas-Fermi Approximation (TFA), which ignores the kinetic energy
terms in the Gross-Pitaevskii equations for the condensate wave
functions \cite{gp}. It has been shown
that in the case of one component condensates the TFA results
agree well  with the numerical
calculations for large particle numbers,
except for a small region near the boundary of the condensate
\cite{tfa1,tfa2}. In fact, even for small numbers of particles
TFA still usually gives qualitatively correct results.

In the dimensionless variables, the Gross-Pitaevskii equations for
the condensates in the harmonic traps may be written in the form
\cite{crt1,crt2}:
\begin{eqnarray}
&-&{\nabla'}^2\psi_1'+({x'}^2+{y'}^2+\lambda^2(z'+z_0')^2)\psi_1'-
\nonumber\\
&-&\mu_1'\psi_1'+
u_1|\psi_1'|^2\psi_1'+
\frac{8\pi a_{12}N_2}{a_{\perp}}|\psi_2'|^2\psi_1'=0;
\label{1}\\
&-&\beta^2{\nabla'}^2\psi_2'+({x'}^2+{y'}^2+\lambda^2(z'-z_0')^2)
\psi_2'-\nonumber\\
&-&\mu_2'\psi_2'+
u_2\beta^2|\psi_2'|^2\psi_2'+
\frac{8\pi a_{12}N_1}{a_{\perp}}|\psi_1'|^2\psi_2'=0;
\label{2}
\end{eqnarray}
Here $\psi_i({\bf r})=\sqrt{N_i/a_{\perp}^3}\psi_i'({\bf r}')$,
$\psi_i({\bf r})$ being the wave function of the species $i$ of a
two-species condensate ($i=1,2$). $\lambda=\omega_z/\omega$.
${\bf r}=a_{\perp}{\bf r}'$,
where
$a_{\perp}=\left(\hbar/m\omega\right)^{1/2}$. $\omega$ is the
trapping frequency. $\mu_i'=2\mu_i/\hbar\omega$, where $\mu_i$ is
the chemical potential of the species $i$. The chemical potentials
$\mu_1$ and $\mu_2$ are determined by the relations
$\int\,d^3|\psi_i|^2=N_i$. $u_i$ is given by
$u_i=8\pi a_{ii}N_i/a_{\perp}$.
The wave function $\psi_i'({\bf r}')$ is normalized to $1$.
$z_0'$ denotes the shift of the minimum of the trapping potential
in the vertical direction.

Equations (\ref{1}) and (\ref{2}) were obtained
by minimization of the
energy functional of the trapped bosons  given by:
\begin{eqnarray}
E'&=&\frac{1}{2}\int\,d^3r'\left[N_1|\nabla'\psi_1'|^2+
N_2\beta^2|\nabla'\psi_2'|^2+\right.\nonumber\\
&+&N_1({x'}^2+{y'}^2+
\lambda^2(z'+z_0')^2)|\psi_1'|^2+
\frac{1}{2}N_1u_1|\psi_1'|^4+\nonumber\\
&+&N_2({x'}^2+{y'}^2+\lambda^2(z'-z_0')^2)|\psi_2'|^2+
\nonumber\\
&+&\left.\frac{1}{2}N_2u_2\beta^2|\psi_2'|^4+
\frac{4\pi a_{12}}{a_{\perp}}
N_1N_2|\psi_1'|^2|\psi_2'|^2\right].
\label{en}
\end{eqnarray}
Here the energy of the system $E$ is related to $E'$ by $E=\hbar\omega
E'$

In the TFA, Eqs. (\ref{1}),  (\ref{2}) and (\ref{en}) can be further
simplified by omitting the kinetic energy terms.
In the framework of TFA the phase segregated condensates do not overlap,
so we can neglect
the last terms in Eqs. (\ref{1}), (\ref{2}) and (\ref{en}), obtaining
simple algebraic equations:
\begin{eqnarray}
|\psi_1'({\bf r}')|^2&=&\frac{1}{u_1}\left(\mu_1'-({r'}^2+
\lambda^2(z'+z_0')^2)\right)\times \nonumber\\
&\times&\Theta\left(\mu_1'-({r'}^2+\lambda^2(z'+z_0')^2)\right)\times \nonumber\\
&\times&\Theta\left({r'}^2+\lambda^2(z'-z_0')^2-\mu_2'\right); \label{3} \\
|\psi_2'({\bf r}')|^2&=&\frac{1}{u_2}
\left(\mu_2'-({r'}^2+\lambda^2(z'-z_0')^2)\right)\times \nonumber\\
&\times&\Theta\left(\mu_2'-({r'}^2+\lambda^2(z'-z_0')^2)\right)\times \nonumber\\
&\times&\Theta\left({r'}^2+\lambda^2(z'+z_0')^2-\mu_1'\right). \label{4}
\end{eqnarray}
Here $\Theta$ denotes the unit step function and ${\rho'}^2={x'}^2+{y'}^2$.
If $z_0'=0$, from Eqs. (\ref{3}) and (\ref{4}) one can see
that the condensate density has the ellipsoidal form.
This case has
been considered in detail in Refs. \cite{crt1,crt2}.

In the case of phase separation, the energy of the system can be written
in the form \cite{crt1,crt2} $E=E_1+E_2$,
where
\begin{eqnarray}
E_1&=&\frac{1}{2}\hbar\omega N_1\left[\mu_1'-
\frac{1}{2}u_1\int\,d^3r'|\psi_1'|^4\right], \label{e1}\\
E_2&=&\frac{1}{2}\hbar\omega N_2\left[\mu_2'-\frac{1}{2}u_2
\int\,d^3r' |\psi_2'|^4\right]. \label{e2}
\end{eqnarray}

To determine the position of the boundary between
the condensates, we use the condition of thermodynamic equilibrium
\cite{landau}: the pressures exerted by both condensates must be equal:
$ P_1=P_2.$ Pressure is given by \cite{pitaevskii}:
$P_i=G_{ii}|\psi_i|^4/2,$ where $G_{ii}=4\pi\hbar^2a_{ii}/m_i.$
Using these equations one can obtain the equation
for the phase boundary:
\begin{equation}
{r''}^2+\left(\lambda z''-\frac{\alpha(\kappa+1)}
{\kappa-1}\right)^2=R^2,
\label{5}
\end{equation}
where $z'=\sqrt{\mu_1'}z'',  r'=\sqrt{\mu_1'}r'', {r'}^2={x'}^2+{y'}^2,
\alpha=\lambda z_0''$, $\kappa=\sqrt{a_{11}/a_{22}}$, and
\begin{equation}
R^2=\frac{\mu_1'-\kappa\mu_2'}{\mu_1'(1-\kappa)}+
\frac{4\alpha^2\kappa}{(\kappa-1)^2}. \label{6}
\end{equation}

From Eqs. (\ref{5}) and (\ref{6}) one can easily understand why
the small displacement of the centers of the trapping potentials
leads to the significant resulting separation of the centers of
mass of the condensates \cite{dhall}.
The basic physics of this amplification of the trap center difference
comes from the two possible final configurations
of the mixture and that the system is
$\bf close$ to the ``critical point'' that separates
the two final configurations.
The two configurations are the symmetric one where one component is inside
and the other component is outside and the asymmetric
one\cite{ohb98,chui99,ao98} where the two
components are on opposite sides. The former configuration is favored when
$\kappa=\sqrt{a_{11}/a_{22}}$ is different from one, with the less repulsive
component in the middle where the density is higher. The asymetric
configuration possess a lower interface energy and is favored when $\kappa$
is close to one. We found that in the Thomas-Fermi approximation,
when the trapping frequencies for the two components are the same,
the amplification factor is proportional to $1/(\kappa-1)$.

From Eqs. (\ref{5})-(\ref{6}) the evolution of the system upon increasing
$\alpha$ may be described as follows: for $\alpha=0$ condensate 1 forms
the shell about the ellipsoidal condensate 2. The semiaxis of
this ellipsoid
is given by Eq. (\ref{5}) for $\alpha=0$. Upon increasing $\alpha$
the inner ellipsoid
moves upwards, while external one moves down. It may be shown that they
touch each other for the critical value of $\alpha$:
$\alpha_c=\frac{1}{2}\left(1-\sqrt{\frac{\mu_2'}{\mu_1'}}\right)$.
For $\alpha>\alpha_c$ phase boundary (\ref{5}) intersect boundaries of
condensates at the points with coordinates:
\begin{eqnarray}
\lambda z_c''&=&\frac{\alpha}{\kappa-1}-
\frac{(\kappa-1)(R^2-1)}{4\alpha\kappa}, \label{8}\\
r_{1,2}''&=&\pm\sqrt{1-(\lambda z_c''+\alpha)^2}, \label{9}
\end{eqnarray}
which can be obtained from Eqs. (\ref{3})-(\ref{5}).
Critical value $\alpha_c$ is a function of the ratio $N_2/N_1$.

Using normalization condition $\int |\psi'_i({\bf r'})|^2\,d^3r'=1$,
one can determine the chemical potentials $\mu'_i$ as functions of
$N_1$, $N_2$, and $\alpha$. Analytical expressions for $\mu_i'$
are different for $\alpha<\alpha_c$ and for $\alpha>\alpha_c$. In
the former case one has:
\begin{eqnarray}
\frac{(\mu_1')^{(5/2)}}{(\mu_1^0)^{(5/2)}}&=&\frac{1}{1-
\frac{5}{2}R^3[1-\gamma^2]+\frac{3}{2}R^5},\label{mu1}\\
\frac{(\mu_1')^{(5/2)}}{(\mu_2^0)^{(5/2)}}&=&\frac{2}{15}\frac{1}
{\frac{R^3}{3}[\frac{\mu_2'}{\mu_1'}-\tilde{\kappa}^2]-\frac{R^5}{5}},
\label{mu2}
\end{eqnarray}
where $\mu_i^0=\left(\frac{15\lambda u_i}{8\pi}\right)^{2/5},
\gamma=\frac{2\alpha\kappa}{\kappa-1}, \tilde{\kappa}=\gamma/\kappa.$
In the limit $\alpha\rightarrow 0$ one has the results
obtained in our previous papers for the non-displaced potential
\cite{crt1,crt2}.
In the case $\alpha>\alpha_c$ the formulas for $\mu_1'$ and $\mu_2'$
obtained after tedious but straightforward calculations are rather
cumbersome and will be given elsewhere. In this article we
discuss the results of calculations. To be specific, we will use
the parameters corresponding to the  experiments on $^{87}Rb$ atoms
$a_{\perp}=2.4\times 10^{-4} cm$,
$N=N_1+N_2=0.5\times10^6$ atoms, $\lambda=\sqrt{8}$.

In Fig.1 we show the density profiles of the condensates (see Eqs.
(\ref{3}) and (\ref{4})) as functions of the vertical
coordinate $z$ for ${r'}^2=0$ and $N_1=N_2$. In this case
$\alpha_c=0.0047$. Fig.1(a) illustrates the experimental situation
\cite{dhall}: $\alpha=0.03$ (approximately 3\% of the extent
of the density distribution in the vertical direction) is larger
than the critical value $\alpha_c$, and condensates are completely
separated in vertical direction in accordance with the experiment
\cite{dhall}. In the case $\alpha<\alpha_c$ the condensate $N_2$
is inside the condensate $N_1$.
It should be noted that rather small shifts of the
trapping potential centers
with respect to each other produce considerable displacements of
the condensates. The condensates in Fig.1 does not overlap
because, as was mentioned earlier,
in the framework of TFA it is impossible to describe
the overlapping of the condensates.

Another interesting question is how the vortex states change when
the minima of the trapping potentials $V_1$ and $V_2$ are
displaced with respect to each other. In a frame rotating with the
angular velocity $\Omega$ along the $z$-axis
the energy functional of the system is:
\begin{eqnarray}
E_{rot}(l_1,l_2)&=&E(\psi_{l_{1}},\psi_{l_{2}})+\nonumber\\
&+&\int\,d^3 r
(\psi_{l_{1}}^*+\psi_{l_{2}}^*)i\hbar\Omega \partial_\phi
(\psi_{l_{1}}+\psi_{l_{2}}), \label{11}
\end{eqnarray}
where $\psi_{l_{j}}({\bf r})=|\psi_{l_{j}}({\bf r})|e^{il_j\phi}$
is the wave function for the vortex excitation with angular
momentum $\hbar l_j$.
In the TFA, the vortex induced change in condensate density is
negligible \cite{fetter} (hydrodynamic approximation).

In the case of the phase segregated condensate, one finds
from Eqs. (\ref{11}) and (\ref{e1}-\ref{e2}) that the energy
change due to presence of the vortices
$\Delta E=E_{rot}(l_1,l_2)-E_{rot}(0,0)$ has the form \cite{crt1,crt2}:
\begin{eqnarray}
\Delta E&=&\Delta E_{N_1}+\Delta E_{N_2}=\nonumber\\
&=&\frac{1}{2}\hbar \omega N_1 \int\,d^3 r'
\left(\frac{l_1^2}{{\rho'}^2}|\psi_1'|^2-
\frac{2\Omega l_1}{\omega}|\psi_1'|^2
\right)+\nonumber\\
&+&
\frac{1}{2}\hbar \omega N_2 \int\,d^3 r'
\left(\frac{l_2^2}{{\rho'}^2}|\psi_2'|^2-\frac{2\Omega l_2}{\omega}
|\psi_2'|^2 \right). \label{12}
\end{eqnarray}
In the hydrodynamic limit $\psi_i'$ is given by Eqs. (\ref{3}) and
(\ref{4}).

In the case $\alpha=0$ critical velocities as functions of ratio
$N_2/N_1$ have been calculated in \cite{crt1}. It was shown that
for all values of $N_2/N_1$, the critical velocity $\Omega_{N_2}$
of the inner condensate is lower than the critical velocity
$\Omega_{N_1}$ of the outer one. So upon increasing $\Omega$, a
vortex will appear first in the external condensate. However, if
for given $\Omega$ one shifts the centers of the trapping
potentials with respect to each other in the vertical direction,
the inner condensate floats to the surface. In this case one
can expect that the critical velocities of the condensates become
closer and even can be equal for some values of $\alpha$ and
$N_2/N_1$.

As in the case of the chemical potentials,
the expressions for the critical
velocities have different analytical forms for $\alpha<\alpha_c$
and for $\alpha>\alpha_c$, the latter being rather cambersome. For
$\alpha<\alpha_c$ the critical velocities are given by:
\begin{eqnarray}
\frac{\Omega_{N_1}}{\omega}&=&\frac{5l_1(\mu_1')^{(3/2)}}{2(\mu_1^0)^{(5/2)}}
\left\{\left(\ln\frac{2\mu_1'}{l_1}-\frac{4}{3}\right)\right.-\nonumber\\
&-&\frac{3}{2}R\left[\left(1-\gamma^2-\frac{R^2}{3}\right)
\ln\frac{2R\mu_1'}{l_1}-\right.\nonumber\\
&-&\left.\left.\left(1-\gamma^2-\frac{R^2}{9}\right)\right]\right\},
\label{om1}\\
\frac{\Omega_{N_2}}{\omega}&=&\frac{15l_2(\mu_1')^{(3/2)}R}{4(\mu_2^0)^{(5/2)}}
\left[\left(\frac{\mu_2'}{\mu_1'}-\tilde{\kappa}^2-\frac{R^2}{3}\right)
\times\right.\nonumber\\
&\times&\left.\ln\frac{2R\sqrt{\mu_1'\mu_2'}}{l_2}-
\left(\frac{\mu_2'}{\mu_1'}-\tilde{\kappa}^2-\frac{R^2}{9}\right)\right].
\label{om2}
\end{eqnarray}
Again, in the limit $\alpha\rightarrow 0$ one has the results
obtained in our previous paper for the non-displaced potential
\cite{crt1}.

Figure 2 shows the behaviour of critical velocities as
functions of $\alpha$ for different values of $N_2/N_1$. Dashed
lines correspond to the  inner condensate, solid lines - to outer
one. From Fig.2(c) one can see that the critical velocities really
can intersect. Physically this means that there is a possibility
to exchange the vortex states between condensates by shifting the
centers of the trapping potentials with respect to each other for
fixed angular velocities.

In summary, we investigated the behaviour of simultaneously
trapped codensates consisting of $^{87}Rb$ atoms in two different
hyperfine states. It is shown that the small shift of the minima
of the trapping potentials with respect to each other leads to the
considerable displacement of the centers of mass of the
condensates in agreement with the experiment \cite{dhall}. It is
also shown that the critical angular velocities of the vortex
states of the condensates drastically depend on the shift and
relative number of particles in the condensates. The predicted
exchange of the vortex states between the condensates as a
function of the shift remains to be studied experimentally.

This work is supported in part by NATO Grant No. PST.CLG.976038.
V.N.R. and E.E.T. are grateful to Bartyol Research Institute of
the University of Delaware for hospitality.

{\bf Figures captions}

Fig.1.Density profiles of the condensates as functions of the
vertical coordinate $z$ for $N_1=N_2$. Figure (a) corresponds to
$\alpha>\alpha_c$ and figures (b)-(c) -- to $\alpha<\alpha_c$.
Solid lines correspond
to the $|1>$ atoms and dashed lines - to the $|2>$ atoms.

Fig.2. Critical velocities of outer condensate $\Omega_{N_1}/\omega$ and
the inner condensate $\Omega_{N_2}/\omega$ as
functions of $\alpha$ for different values of $N_2/N_1$. Dashed
lines correspond to $\Omega_{N_2}/\omega$ and solid lines -
to $\Omega_{N_1}/\omega$.


\begin{thebibliography}{99}

\bibitem{[1]}  M.\,N. Anderson, J.\,R. Ensher, M.\,R. Matthews
{\it et al.}, Science {\bf 269}, 198
(1995).
\bibitem{[2]} K.B. Davis {\it et al.}, Phys. Rev. Lett. {\bf 75},
3969 (1995)
\bibitem{[3]} C.\,C. Bradley, C.\,A. Sackett, and R.\,G. Hulet, Phys.
Rev. Lett. {\bf 78}, 985 (1997).
\bibitem{stir2000} K.\,W. Madison, F. Chevy, W. Wohlleben
{\it et al.}, Phys. Rev. Lett. {\bf 84}, 806 (2000).
\bibitem{bose_vor} M.\,R. Matthews, B.\,P. Anderson, P.\,C. Haljan
{\it et al.}, Phys. Rev. Lett. {\bf 83}, 2498 (1999).
\bibitem{wilhol} J.\,E. Williams and M.J. Holland, Nature
{\bf 401}, 568 (1999).
\bibitem{dhall}  D.\,S. Hall, M.\,R. Matthews, J.\,R. Ensher
{\it et al.}, Phys. Rev. Lett. {\bf 81},
1539 (1998).
\bibitem{twocom1} M.\,R. Matthews, D.\,S. Hall, D.\,S. Jin
{\it et al.}, Phys. Rev. Lett.
{\bf 81}, 243 (1998).
\bibitem{twocom2} D.\,S. Hall, M.\,R. Matthews, C.\,E. Wieman
{\it et al.}, Phys. Rev. Lett. {\bf 81},
1543 (1998).
\bibitem{crt1} S.\,T. Chui, V.\,N. Ryzhov, and E.\,E. Tareyeva,
JETP {\bf 91}, 1183 (2000).
\bibitem{crt2} S.\,T. Chui, V.\,N. Ryzhov, and E.\,E. Tareyeva,
Phys. Rev. A {\bf 63}, 023605 (2001).
\bibitem{gp}  L.\,P. Pitaevskii, Sov. Phys. JETP {\bf 13}, 451 (1961);
E.\,P. Gross, Nuovo Cimento {\bf 20}, 454 (1961); J. Math.
Phys. {\bf 4}, 195 (1963).
\bibitem{tfa1}  G. Baym and C.\,J. Pethick, Phys. Rev. Lett.
{\bf 76}, 6 (1996).
\bibitem{tfa2}  F. Dalfovo and S. Stringari, Phys. Rev. A {\bf 53},
2477 (1996).
\bibitem{landau} L.\,D. Landau and E.\,M. Lifshits. {\it Statistical Physics.
Part I,} 3rd ed. (Nauka, Moscow, 1976).
\bibitem{pitaevskii} L.\,P. Pitaevskii, Usp. Fiz. Nauk, {\bf 168}, 641 (1998)
[Phys. Uspekhi {\bf 41}, 569 (1998)].
\bibitem{ohb98}  P. Ohberg and S. Stenholm, {\it Phys. Rev. A} {\bf 57},
1272 (1998).
\bibitem{chui99}  S.\,T. Chui and P. Ao, {\it Phys. Rev. A }{\bf 59}, 1473
(1999).
\bibitem{ao98}  P. Ao and S.\,T. Chui, {\it Phys. Rev. A }{\bf 58}, 4836
(1998).
\bibitem{fetter}  A.\,L. Fetter, cond-mat/9811366
(published in {\it Bose-Einstein
Condensation in Atomis Gases}, Proceedings of the International
School of Physics "Enrico Fermi", ed. M. Inguscio, S. Stringari,
and C. Wieman (IOS Press, Amsterdam, 1999)).

\end{thebibliography}
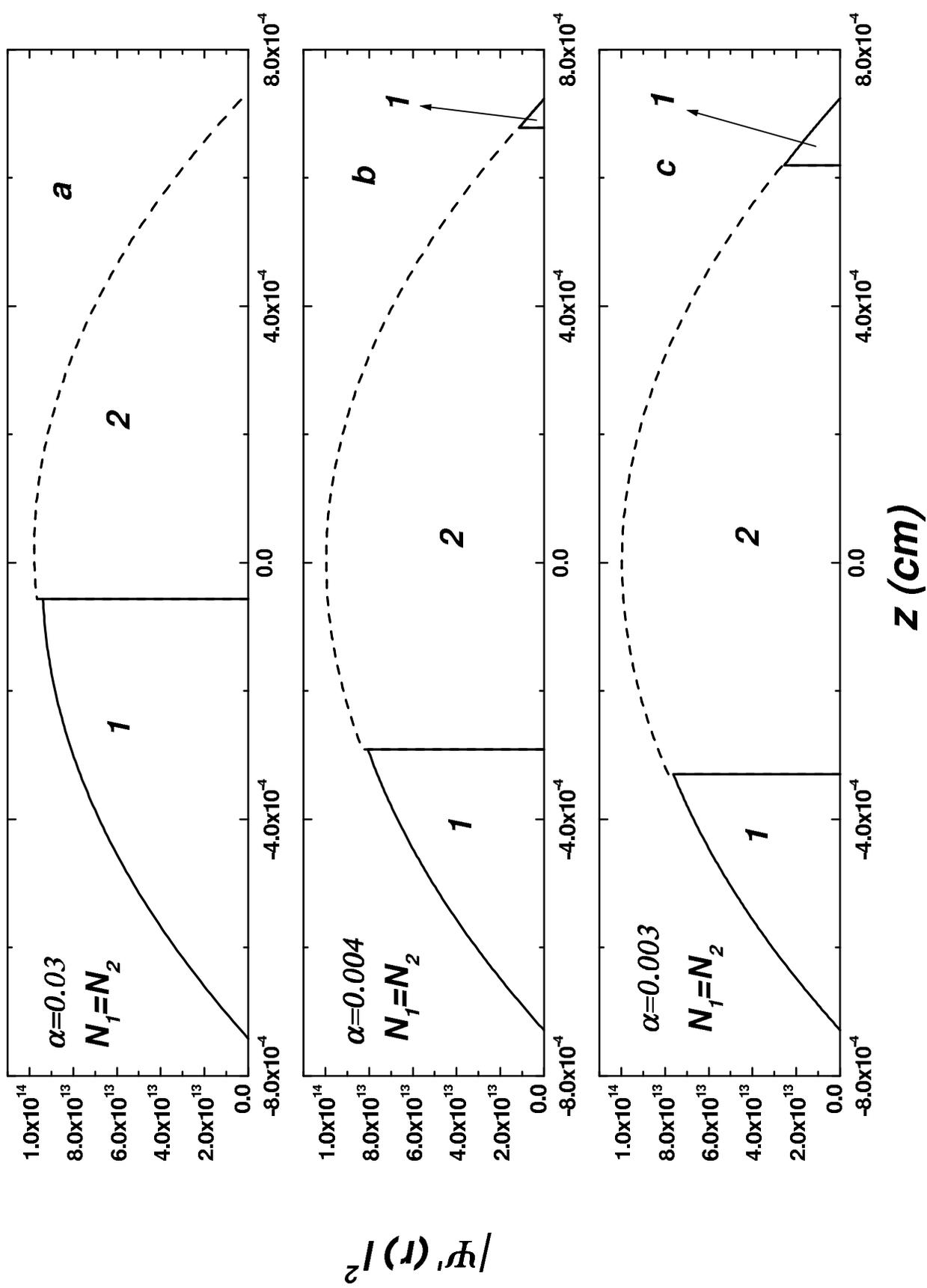
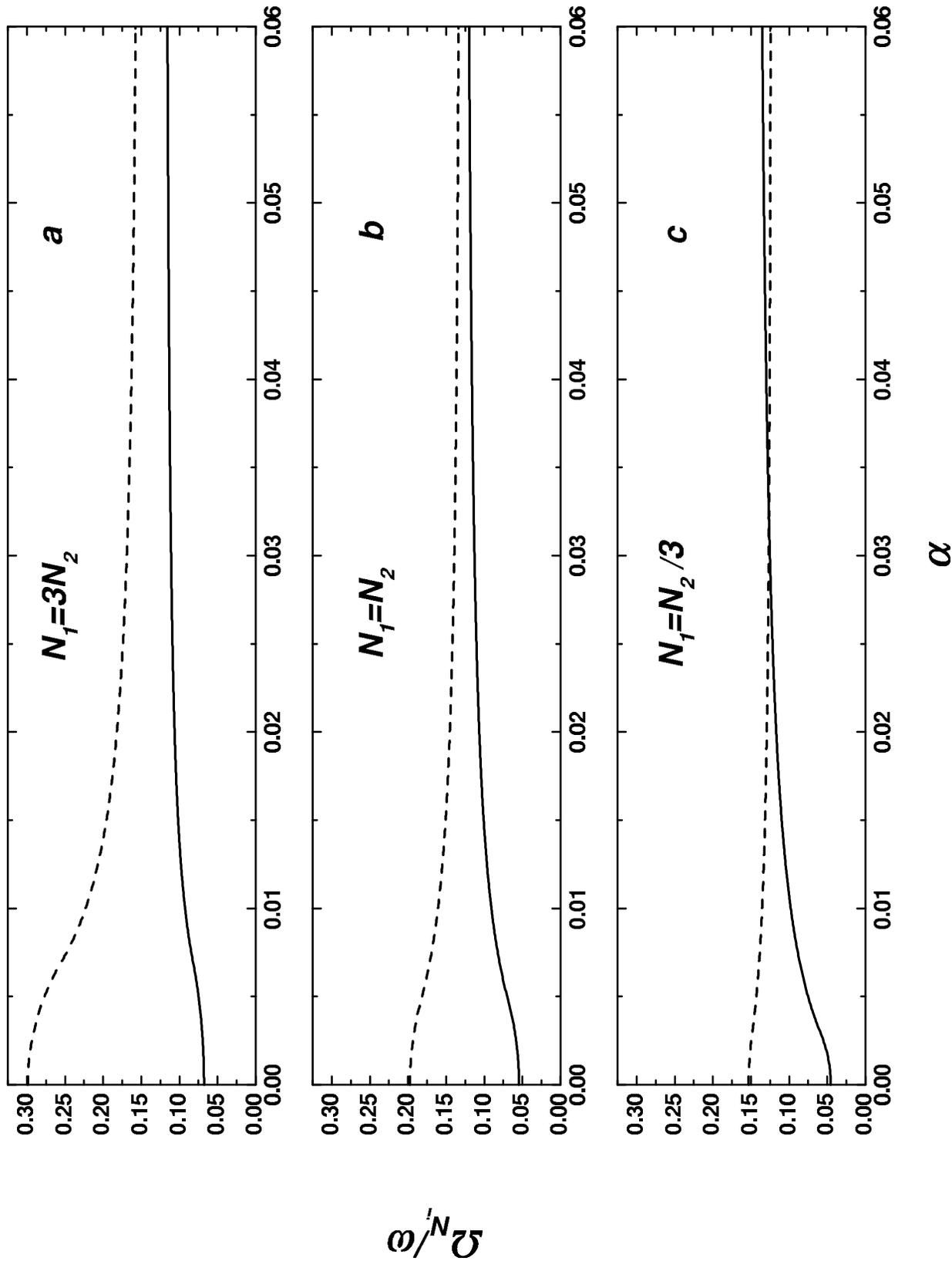
\end{document}